\documentclass[12pt]{article}
\usepackage{graphicx,psfrag,epsf}
\usepackage{enumerate}
\usepackage{url} 

\usepackage{amsmath}
\usepackage{amsfonts}
\usepackage{amsthm}
\usepackage{amssymb}
\usepackage{multimedia}
\usepackage{verbatim}
\usepackage{times}
\usepackage[T1]{fontenc}
\usepackage{booktabs}
\usepackage{tabularx}
\usepackage{colortbl}
\usepackage{graphicx}
\usepackage{amsmath}
\usepackage{amsfonts}
\usepackage{amsthm}
\usepackage{fancyhdr}
\usepackage{framed}
\usepackage{natbib}
\usepackage{ulem}  
\usepackage{algorithmic}
\usepackage{algorithm}

\newcommand{\bm}[1]{\mbox{\boldmath $#1$}}
\newcommand{\mb}[1]{\mathbf{#1}}

\newcommand{\blind}{0}

\begin{document}

\def\spacingset#1{\renewcommand{\baselinestretch}%
{#1}\small\normalsize} \spacingset{1}


\if0\blind
{
  \title{\bf Computer experiments with functional inputs and scalar outputs by a norm-based approach}
  \author{Thomas Muehlenstaedt \thanks{
    The authors gratefully acknowledge}\hspace{.2cm}\\
    W. L. Gore \& Associates \\
    and \\
    Jana Fruth  \\
    Faculty of Statistics, TU Dortmund\\
    and \\
    Olivier Roustant\\
    Ecole Nationale Superieure des Mines de Saint-Etienne,\\ FAYOL-EMSE, 
    LSTI, F-42023 Saint-Etienne, France\\}
  \maketitle
} \fi

\if1\blind
{
  \bigskip
  \bigskip
  \bigskip
  \begin{center}
    {\LARGE\bf Computer experiments with functional inputs and scalar outputs by a norm based approach}
\end{center}
  \medskip
} \fi

\bigskip
\begin{abstract}
A framework for designing and analyzing computer experiments is presented, which is constructed for dealing with functional and real number inputs and real number outputs. For designing experiments with both functional and real number inputs a two stage approach is suggested. The first stage consists of constructing a candidate set for each functional input and during the second stage an optimal combination of the found candidate sets and a Latin hypercube for the real number inputs is searched for. The resulting designs can be considered to be generalizations of Latin hypercubes.
GP models are explored as metamodel. The functional inputs are incorporated into the kriging model by applying norms in order to define distances between two functional inputs. In order to make the calculation of these norms computationally feasible, the use of B-splines is promoted.
\end{abstract}

\noindent%
{\it Keywords:}  space-filling design, Gaussian process, Maximin design
\vfill

\newpage
\spacingset{1.45} 

\section{Introduction}\label{sec_intro}
A lot of physical phenomena are now studied virtually by means of computer codes.
For complex phenomena it often happens that the code is too time-consuming for a direct usage.
This issue is usually addressed by creating ``metamodels'', also called ``surrogates'' or ``emulators'',
that correspond to quick-to-evaluate mathematical models of the computer codes. 
In particular the (meta)model based on a Gaussian process (GP) proposed by  
\cite{SacSchWel89desi, SacWelMit89desi} and \cite{CurMitMor91baye} at the end of the eighties
has gained in popularity, and is now described in several books 
(see e.g. \cite{SanWilNot03desi}, \cite{FanLiSud06desi}, \cite{GPML}).
In the sequel we will use the term ``GP model'' though other equivalent expressions can be found, 
such as GP Regression, GaSP, GP emulator, or Kriging model.
One main reason for its success is that the GP model 
both provides an interpolation of the data and an uncertainty quantification in the unexplored regions.
Furthermore, it depends on a positive definite function, or \textit{kernel},
that is adaptable to specific priors or frameworks.

A large amount of research has addressed the case of scalar-valued inputs and outputs,
though this is often a summary of functional inputs and outputs, given as functions of time or space.
Nevertheless there has been a recent literature focusing on this more complex functional framework.
\cite{BayarriEtc2007ValidationFunOut} investigated model validation with functional outputs
by using a wavelet decomposition.
\cite{ShiWanMurTit2007GPFRforBatchData} had batches of time-varying data,
and modeled separately the mean structure with a functional regression model (\cite{RamSil97func})
and the covariance structure with a GP model. 
Developments are given in \cite{ShiShoi2011GPFunctionalDataBook}.
\cite{Morris2012KrigingWithTimeVaryingInputs} introduced a new kernel for the GP model
that allows modelling time-varying inputs and outputs. He also considered the design problem,
and extended the maximin distance to the time-varying case. Some theoretical results on designs for computer experiments with time varying inputs can be found in \cite{Morris2014MaximinDoE}. 

In this article we consider the situation where the inputs are either scalar-valued or functional,
and where the output is scalar-valued. This corresponds, for instance, 
to practical situations where engineers study a summary of the output, 
but consider the whole complexity of the inputs,
that may be scalar-valued but also time-varying functions or more general multivariate functions.
We investigate a GP model approach using a customized kernel based on norms and B-splines.
We also propose an original design strategy aiming at providing an initial space-filling design. Although the methods are related to the work presented by \cite{Morris2012KrigingWithTimeVaryingInputs}, it covers different cases, e.g. our method is not restricted to time varying inputs and, at the same time, time varying outputs. Furthermore we allow for a combination functional and scalar inputs. 

Section 2 provides some notations and presents the functional framework, 
including some basics about B-splines and functional norms. 
In Section 3 designs for computer experiments with both functional and scalar-valued inputs are described. 
In Section 4, GP models are derived, including a weighting procedure 
for extracting which part of a functional input has high influence on the output. 
In Section 5, the methodology is applied to a theoretical example and to a sheet metal forming problem.
A concluding discussion is given in Section 5.

\section{Background and notations}
In this paper we consider a scalar-valued function $g$
depending on functional inputs $\mb{f(t)} = (f_1(t), \dots,  f_{d_f}(t) )$,
as well as, possibly, on scalar-valued inputs $\mb{x} = (x_1, \dots, x_{d_s})$:

\begin{equation}
y = g(\mb{x}, \mb{f}(t)) 
\end{equation}

\noindent In the notations above, $g$ represents a time-consuming simulator
and $d_s, d_f$ are two integers, with $d_f  > 0$. 
For the sake of simplicity we consider that $t \in [0,1]$ is scalar-valued 
but the methodology presented here could be generalized to a vector-valued input $\mb{t} \in [0, 1]^{d_t}$.
We assume that the inputs are bounded, and have been rescaled to $[0,1]$: 
$\mb{x} \in [0, 1]^{d_s}$ and $f_j(t) \in [0,1]$ for all $t \in [0,1]$ and $j \in \{1, \dots, d_f\}$.
In this framework, a design of experiments $\mathcal{D}$ with $n$ runs consists of  
two sets of scalar and functional inputs denoted by $\mb{x}^{(i)} = (x^{(i)}_1, \dots, x^{(i)}_{d_s})$
and $\mb{f}^{(i)}(t) = (f^{(i)}_1(t), \dots, f^{(i)}_{d_f}(t))$, $i=1,\dots,n$.
We denote by $y^{(1)}, \dots, y^{(n)}$ the corresponding scalar-valued outputs.
The design used is denoted by $\mathcal{D}$, in contrast to a distance later on denoted by $D$.
\subsection{Some basics on B-splines}
\label{sec:splines}

B-splines  are an attractive tool for the modeling of functional input (see \cite{deBoor2001GuideToSplines}, \cite{RamSil97func}). They cover various types of functions, reduce the infinite space of
functions considerably and provide a practical mathematical framework for further
computations. B-spline functions are always bounded, which is an important feature for input
functions which usually are only allowed to vary between given values.

A B-spline is defined as a linear combination of B-spline basis functions $B_{i,m}, i=1,\dots,K$ of order $m$
\[
f(t) = \sum_{i=1}^K \beta_i B_{i,m}(t)
\]
where the order $m=1$ relates to (piecewise) constant functions. $K$ and $m$ have to be fixed with $K\geq m$ and
$\bm{\beta}=(\beta_1,\dots,\beta_K)$ is the vector of basis coefficients. The B-spline basis functions are
defined over a sequence of increasing knots (time points) of length $K-m+2$ with additional
$m-1$ replicates for the first and the last knot which are necessary for basis functions at the
bounds
\[
\tau_1= \dots =\tau_{m-1} = \tau_m < \tau_{m+1}< \dots< \tau_{K} < \tau_{K+1}= \tau_{K+2}= \dots= \tau_{K+m}.
\]
They are recursively given by
\[
B_{i,1}(t) = \bm{1}_{[\tau_i, \tau_{i+1}]}(t) \
\]
for $i = 1\dots, K+m-1$ and
\[
B_{i,m}(t) = \frac{t-\tau_i}{\tau_{i+m-1}-\tau_i}B_{i,m-1}(t) + \frac{\tau_{i+m}-t}{\tau_{i+m}-\tau_{i+1}}B_{i+1,m-1}(t)
\]
for $i \in 1, \dots, K$, with $B_{i,m} = 0$ if $\tau_i= \dots = \tau_{i+m} = 0$ to avoid division by zero.

\begin{figure}[ht]
  \centering
  \includegraphics[width=\textwidth]{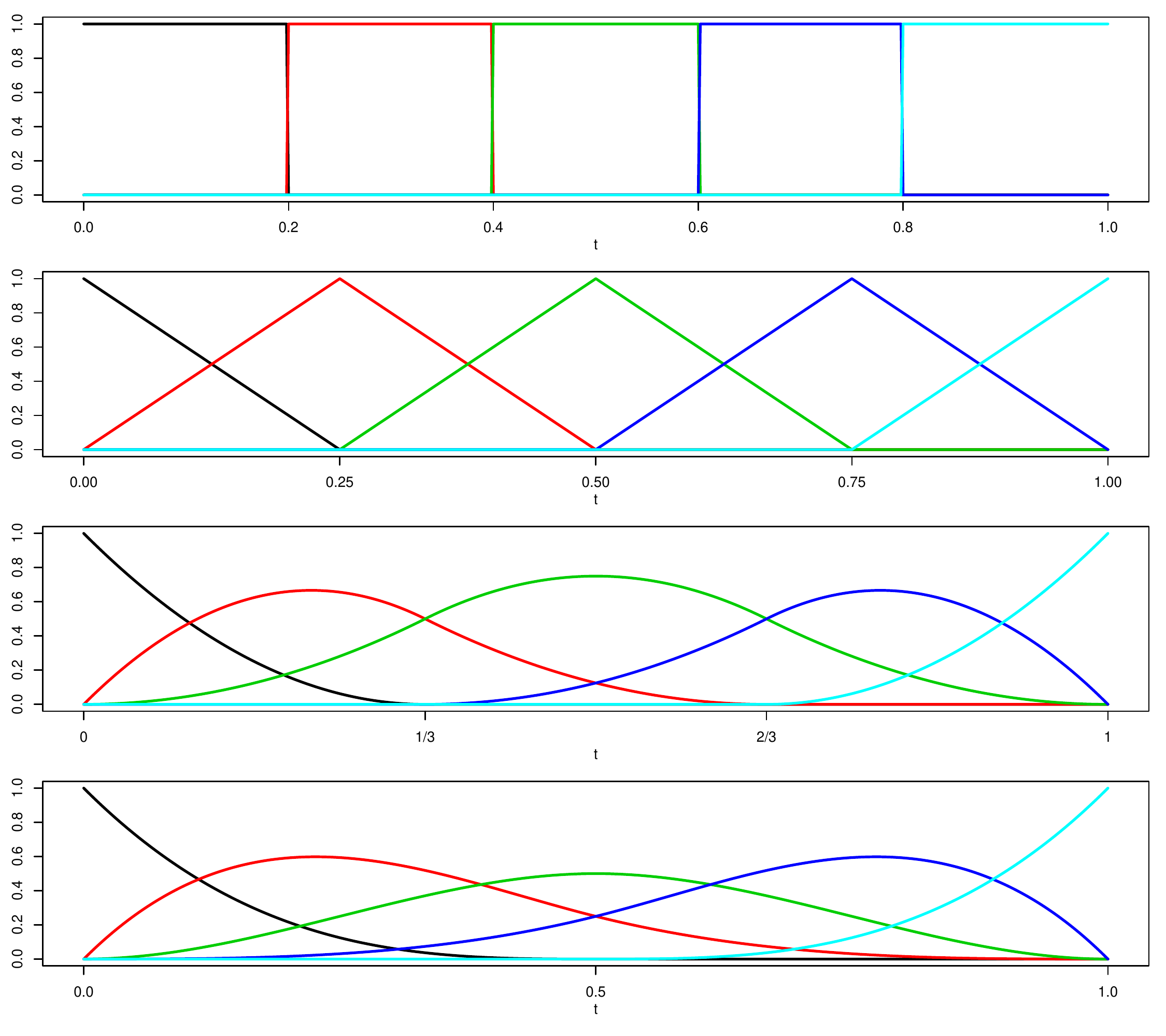}
\caption{B-Spline bases of increasing orders (1 to 4 from top to bottom) for a fix number of
    $K=5$ functions. The knots are shown as ticks on the x-axis.}
\label{fig:spline}
\end{figure}

Figure \ref{fig:spline} shows basis functions for B-splines of order 1, 2, 3, and 4. For order~1 the
basis functions are disjoint piecewise constant functions, for order 4 the functions form the popular cubic spline. In the
figure the number of basis functions $K$ is set to 5 for each order. It follows that the number of knots decreases with
the order. It can be further seen that at each time point $t$, the sum of all 5 basis functions is 1, which
implies that if $\bm{\beta} \in [0,1]^K$ then for all $t$ we have $0 \leq f(t) \leq
1$, $f(t)=1 \Leftrightarrow \beta_1 = \dots = \beta_K = 1$ and $f(t)=0 \Leftrightarrow \beta_{1}=\dots=\beta_{k}=0$. Therefore a bounded function $f$
corresponds to a hypercubic domain for the basis coefficients $\beta$. 

One result which justifies the use of B-Splines is found in \cite{deBoor2001GuideToSplines} on page 55, equation 12. In our notation the result stats that given an unknown but four times differentiable function $g(t)$ defined on $[\tau_m, \tau_{K}]$, the (pointwise) interpolation error of a cubic B-spline is bounded from above and the bound depends on the maximum stepwidth of the knots and the absolute maximum of 4th derivative of the function $g$. 
Hence, while B-splines itself are a somewhat restriced class of functions, they can be used to approximate a very broad class of functions, i.e. all sufficiently smooth functions.

\subsection{Distance-based approach}

To discriminate between functions, a distance based approach is
chosen. Similar to the $L_2$ norm in Euclidean space, the well known $L_2$ norm for functions is
defined as

\begin{equation}\label{eq:d2}
D_f(f, \tilde{f}) = \|f-\tilde{f}\|_{L^2} = \sqrt{\int_0^1 (f(t) - \tilde{f}(t))^2dt}.
\end{equation}

Other choices of norms could be possible, e.g. weighted norms, general $p$ norms or norms working on
derivatives, e.g. Sobolev norms. The choice of a suitable norm is case sensitive. In
the case that the functions are designed as B-splines of the same basis (same number of basis functions, basis order and knot points) the $L^2$ norm of $h(t) = f(t) - \tilde{f}(t) =
\sum_{i=1}^{K} (\beta_i - \tilde{\beta}) b_i(t) = \sum_{i=1}^{K} \delta_i b_i(t)$ reduces to a norm in $\mathbb{R}^{K}$

  \begin{align} \label{eq:Jmatrix}
 \|h\|_{L^2}^2 = &\int_0^1 h(t)^2\,dt \\
 = & \int_0^1 \sum_{i,j}\delta_i\delta_jb_i(t)b_j(t)\,dt\\
 = & \bm{\delta}' J \bm{\delta}\\
= & \|\bm{\delta}\|^2_{J} 
  \end{align}
where $J$ is the $K\times K$ matrix $\left(\int_0^1 b_i(t)b_j(t)\,dt\right)_{1\leq i,j \leq K}$. As the matrix $J$ does not depend on the coefficients of a B-spline function but just on the order and number of basis functions, this matrix can be stored and reused.\\

In order to include the scalar valued inputs into the framework, a further distance function has also to be defined:
\begin{equation}\label{eq:Dgeneral}
    D((x^{(i)}, f^{(i)}), (x^{(j)}, f^{(j)})) = \sqrt{\|x^{(i)} - x^{(j)}\|_2^2 + \sum_{k = 1}^{d_f} (D_f(f_k^{(i)}, f_k^{(j)}))^2}.
\end{equation}
In the case that there are no scalar inputs, this distance simplifies to $\sqrt{\sum_{k = 1}^{d_f} (D_f(f_k^{(i)}, f_k^{(j)}))^2}$.

\section{Designs for functional inputs}

\subsection{Theory}

There are many approaches on how to design a simulation experiment. A good summary can be found in \cite{FanLiSud06desi}. Uniform design criteria like the Wrap Around Discrepancy or the Centered Discrepancy are popular approaches, as well as distance-based design
criteria like maximin and minimax designs. In contrast to uniform and distance-based designs, which
are not directly linked to a statistical model, maximum entropy designs and IMSE optimal are
optimality criteria, which are directly linked to a GP model and an assumed covariance
kernel. A popular class of designs are Latin Hypercube designs (LHD), invented by \cite{MckBecCon79Comp}.

Our aim is to generalize the concept of LHD to situations with functional
inputs. One approach would be to design the coefficients of a basis, such as a B-spline basis,
a polynomial basis, etc. (see \citet{RamSil97func}).

Here another approach is taken. For a
conventional LHD with just scalar inputs, the values of each input variable $x_k$ are
equally spread between $0$ and $1$, i.e. $x_k^{(i)} = \frac{\pi(i) - 1}{n - 1}, i = 1,\dots,n$, 
 where $\pi(.)$ is a permutation of $1, \dots, n$. While it is not obvious, which combination of
the input variables $x_1$ to $x_d$ to use, it is ensured, that the one dimensional projections are
uniformly distributed. The combination is then chosen according to a fitness criterion. This idea is copied
to the functional inputs such that in a first step, a candidate set of functions $f^{(1)}_k, \dots, f^{(n)}_k$
is constructed for each $k \in \{1, \dots, d_f\}$. Once these sets are constructed, the best
combination of the sets and the scalar inputs is determined.

In the following, the strategy for finding a candidate set based on B-splines is described. This corresponds to finding equally spread points in one dimension for conventional LHD with scalar inputs. 
Depending on the restrictions on the candidate set, different strategies for finding a good
candidate set can be applied. If no prior knowledge is available, our strategy is to apply distance-based approaches here as well. After finding a candidate set, in a second step a space-filling combination of the scalar LHD and the candidate set is found.

\subsubsection{Constructing the candidate set}

For B-splines, the choice of the candidate set reduces to the choice of the coefficients of basis
functions, i.e. for a candidate set with $n$ functions with $K$ bases, $n*K$ coefficients have to be
chosen.

In order to have a space-filling candidate set, the coefficients have to be chosen with care. Ideally a big variety of functions are covered, i.e. increasing/decreasing functions or functions which are in average very high or low.

Here the basis coefficients are sampled from a LHD, i.e. each basis function is considered as an input factor in a LHD. 
The coefficients could also be sampled and optimized without any restriction to a LHD, but in this case, the coefficients tend to be near to the extremes for DoEs with larger number of basis functions and higher number of runs.
As a fitness criterion, not directly the minimum distance among all pairs of
functions is used, but a variant proposed in \cite{MorMit95expl}:
\begin{equation}
    \Phi_q(\mathcal{D}_f(\beta)) = (\sum^n_{i = 1} \sum_{j = 1}^{i - 1} (D_f(f^{(i)}(\beta_i),
    f^{(j)}(\beta_j)))^{-q})^{1/q}.
\end{equation}
Here, $q = 5$ is applied. The criterion $\Phi_q(\mathcal{D}_f(\beta))$ is written in dependence on $\beta$, as for the construction of the candidate set the optimization takes place over the coefficient vector of the B-spline representation. This fitness criterion does not only use the minimum distance for
comparison of different designs but all possible pairs of functions. In order to optimize the $\Phi_q$-criterion any existing algorithm for optimizing LHD can be used, e.g. simulated annealing or genetic algorithms. Here, simulated annealing has been used.
\subsubsection{Constructing a generalized Latin hypercube}
Given for each functional input $f_k$ a set of functions is created, the best combination of the
LHD for the real inputs and the sets for the functional input has to be searched. Here the same
set is used for all functional inputs. However, it would be possible to use a different set for each
functional input. In order to rank full designs again a maximin strategy will be chosen.
Therefore the distance (\ref{eq:Dgeneral}) is used for the following criterion:
\begin{equation}
    \Phi^c_q(\mathcal{D}(\pi)) = \left(\sum^n_{i = 1} \sum_{j = 1}^{i-1} (D((x^{(i)}(\pi_i), f^{(i)}(\pi_i)), (x^{(j)}(\pi_j), f^{(j)}(\pi_j))))^{-q}\right)^{1/q}
\end{equation}

The criterion $\Phi^c_q(\mathcal{D}(\pi))$ is written in dependence on a permutation $\pi$ in order to indicate, that in this step of the design construction, the optimization only takes place over switching indices of the scalar of functional inputs.
In order to optimize this fitness criterion by an algorithm, there are multiple algorithms
possible. In principle, all algorithms used for optimizing LHDs can be applied here, as the
candidate sets themselves are not changed, just the combination of the candidate sets and the scalar
inputs. Here a variant of simulated annealing as described in \cite{MorMit95expl} is used.

\newtheorem{remark1}{Remark}
\begin{remark1}
Another alternative for finding designs, which seems to be promising in the first place is to apply a searching
algorithm directly on the fitness criterion by optimizing over the class of functions, the results
is that only extremes of the class are chosen. This is similar to maximin designs with just real
inputs: The optimal maximin design without restricting it to be a LHD in $d$ dimensions
with $2^d$ runs is a traditional full factorial design with two levels, which is definitely not a
space-filling design. In order to illustrate this behaviour, a design with 2 functional and 2 scalar inputs, 15 runs and 8 basis functions has been set up. This design has been optimized unconditionally over the coefficients of the functional inputs and the permutation of the scalar inputs. In Figure \ref{fig:remark1} a plot of the B-splines for the first functional input is given. Clearly the distinctive functions cluster at the minimum and maximum of the allowed range.
\begin{figure}[ht]
  \centering
  \includegraphics[width=\textwidth]{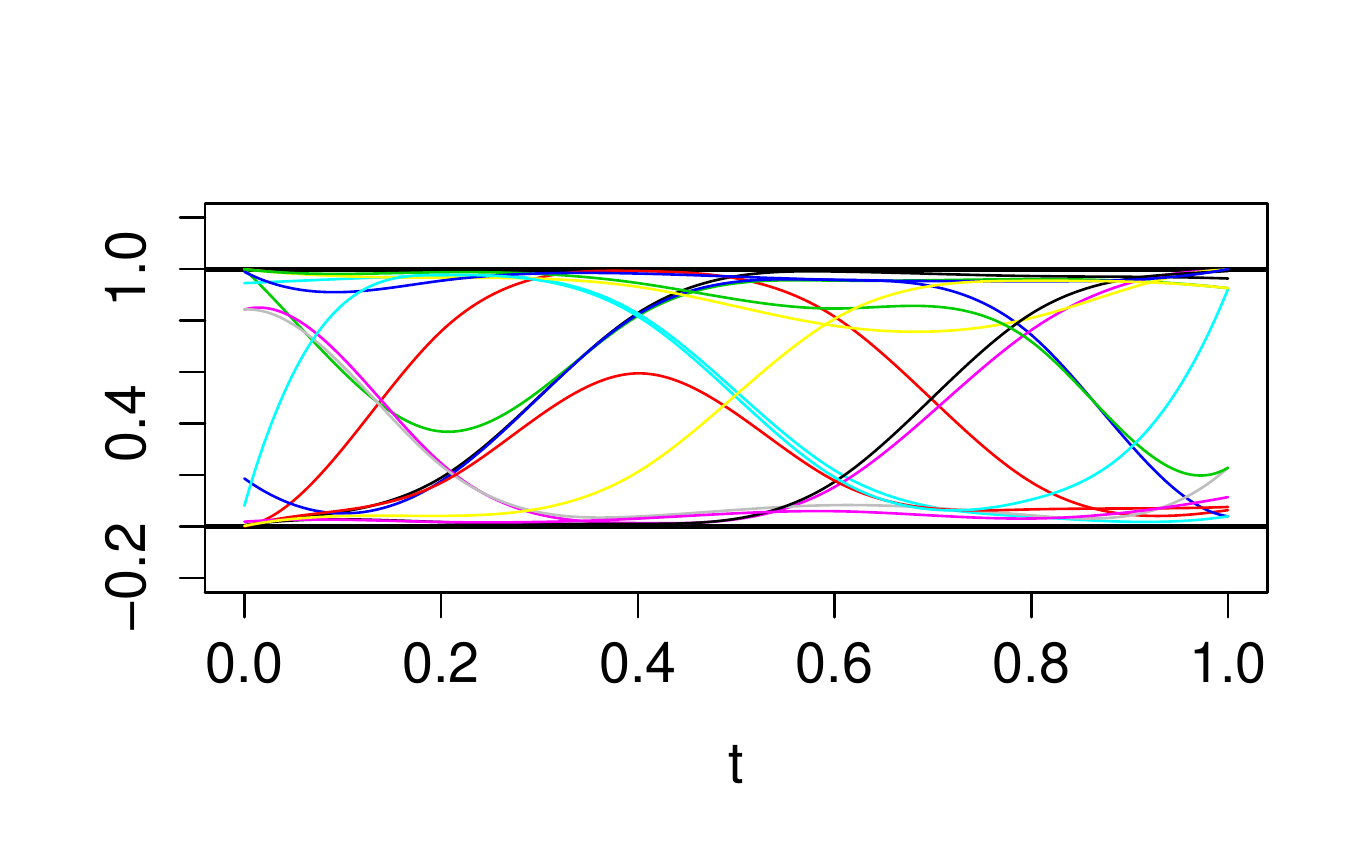}
\caption{Plot of one functional input of an unconditionally optimized design with 15 runs, 2 functional and 2 scalar inputs and 8 basis functions. Clearly, the functions are clustering at the minimum and maximum of the allowed output range.}
\label{fig:remark1}
\end{figure}

\end{remark1}

\section{Surrogate models}

As for many simulations, the evaluation of the simulation is costly, a big part of literature about computer experiments focused on constructing statistical models for simulation output. Different types of models are applied, e.g. response surface models, artifical neural networks or radial  basis functions. However, the most popular model is most likely the GP model (\cite{SanWilNot03desi}), also called Kriging. There are several reasons for using a GP model. It is capable of exactly reproducing the observations, gives an uncertainty estimate and is very flexible by incorporating different covariance kernels. Furthermore, the GP model has often a very high prediction power compared to other approaches and there is an easy way to switch from interpolation to smoothing by incorporating a nugget effect. In \cite{Morris2012KrigingWithTimeVaryingInputs}, a GP model is extended to incorporate time varying inputs, which are modeled as functional inputs. The ideas presented in the following are in some ways extensions of the modeling ideas developed by Morris.

Especially due to the flexibility a GP model is chosen here as well. 
For a standard GP model it is assumed that the output of the simulation follows a Gaussian process:
$$Y(x, f) = \mu + Z(x, f),$$
where the zero centered GP $Z(x, f)$ is characterized by its covariance function. 
A typical GP model approach is to use an anisotropic, tensor-product kernel, which can easily be extended here:
\begin{equation}
    \text{cov}(Z(x^{(1)}, f^{(1)}), Z(x^{(2)}, f^{(2)})) = \sigma^2 g( D_s(x^{(1)},x^{(2)}; \theta_s)) g( D_f(f^{(1)},f^{(2)}; \theta_f)).
\end{equation}

Here $D_f(.,.;\theta_{f})$ and $D_s(.,.;\theta_s)$ are distances for the functional and scalar inputs respectively, scaled by some covariance parameters $\theta_s, \theta_f$.
Standard kernels for $g_k(., \theta)$ are the Gaussian kernel ($g(h, \theta) = exp(\frac{-h^2}{2\theta^2})$), the Matern 5/2 kernel $(g(h, \theta)) = (1 + \frac{\sqrt{5}|h|}{\theta} + \frac{5 h^2}{3 \theta ^2}) exp(-\frac{\sqrt{5}|h|}{\theta})$.

In this statistical model, the parameters $\mu, \sigma, \theta_s$ and $\theta_f$ have to be estimated. 
There are several approaches for parameter estimation in GP models (Maximum likelihood, restricted Maximum Likelihood, cross validation), where the most common is Maximum Likelihood. 
As the likelihood cannot be optimized analytically here algorithmic optimization is chosen.

While in the general case, $\|f-\tilde{f}\|$ requires the evaluation of an integral,
the use of a B-spline basis simplifies the computation. The kernel reduces to a kernel defined on
$D\times D$ where $D$ is the hypercube $[0,1]^{d_s+d_f\times K}$, e.g. a Gaussian covariance kernel reduces
to
\[
\exp \left(- \sum_{\ell=1}^{d_x}\frac{1}{2}\left(\frac{x_\ell^{(1)}-x_\ell^{(2)}}{\theta_{x\ell}}\right)^2\right)
\exp \left(-
  \sum_{\ell=1}^{d_f}\frac{1}{2}\left(\frac{(\bm{\beta}^{(1)}_\ell-\bm{\beta}^{(2)}_\ell)'J(\bm{\beta}^{(1)}_\ell-\bm{\beta}^{(2)}_\ell)}{\theta_{f\ell}}\right)^2\right).
\]
with $f_{\ell} = \sum_{k=1}^K \beta_{\ell,k}B_{k,m}$ for the functional inputs $\ell = 1,\dots,d_f$.
Furthermore the domain is here hypercubic, both for scalar inputs and functional inputs, due to the
property of B-splines (see Section \ref{sec:splines}).

The estimation of the parameters is done in a similar way to the estimation methods used in the $R$ package DiceKriging (\cite{RCoreTeamR}, \cite{RouGinDevDiceKriging}). Therefore in a first step a number of random points in the parameter space are checked for their log-likelihood value and the best is chosen as starting point for the optimization by the $R$-command $optim$.

Many useful concepts, which are known for scalar-valued inputs also work in this context of functional inputs. A leave-one-out cross validation, where the unknown parameters are estimated based on the full data set, but a prediction is made for data point $(x^{(i)}, f^{(i)})$ based on the full data set omitting data point $i$, can be useful to check model adequacy. Although such kind of leave one out prediction is optimistic, it still can help to identify problems with the model.

As for other GP models, an uncertainty estimate is available and hence EGO type optimization techniques (\cite{JonSchonWelch98EGO}) can be applied for sequential optimization.

\subsection{Weighting}

The surrogate model strategy explained above is attractive, as it shrinks down the infinite dimensional functional input to a problem where for each functional input one covariance parameter is estimated. The disadvantage of this approach is that it tells if one functional input as a whole is important or not via its covariance parameter. But it does not give any result about, which part of the input space of a functional input is important. In order to construct a more informative parameter estimation process, a weighting step in the GP model is suggested. So far the distance between two different functions of one functional input is determined by the $L^2$ distance of the two functions and this distance it used as basis for constructing the covariance between two outputs $Y^{(1)}$ and $Y^{(2)}$. For the special case of B-splines, the $L^2$ distance reduces to $\bm{\delta}' J \bm{\delta}$. The general idea for the weighting process is to use instead
\begin{equation}\label{eq:Dtiledf}
    \tilde{D}_w(f, \tilde{f}) = \sqrt{\int_0^1{ (w(t;\omega)* (f(t) - \tilde{f}(t)))^2dt}},
\end{equation}
with $\int_0^1 w(t; \omega)dt = 1$. One of the advantages of using B-splines is that the integration is easily done numerically. As this advantage should not be destroyed, the weighting process has to be chosen carefully.
Writing equation (\ref{eq:Dtiledf}) with $f$ and $\tilde{f}$ being B-splines becomes

\begin{equation}
    \tilde{D}_w(f, \tilde{f}) = \sqrt{\int_0^1{ ( \sum_{i=1}^K \delta_i B_{i,m}(t) w(t;\omega))^2dt}},
\end{equation}
with $\delta_i = \beta_i - \tilde{\beta}_i$
Although this would be in general a possible way for weighting, the numerical computation would be much more complex than before, as the matrix $J$ would now also depend on (weighting) parameters to be estimated. In order to avoid this drawback, the weighting process is discretized such that each basis function is weighted separately:

\begin{equation}
    D_w(f, \tilde{f}) = \sqrt{\int_0^1{ ( \sum_{i=1}^K \delta_i B_{i,m}(t) w_i(\omega))^2dt}},
\end{equation}
with weighting coefficients $w_1(\omega), \dots, w_K(\omega) \geq 0, \sum_{k = 1}^K w_k = 1$.
As the weights can be taken out of the integration, now the integral can again be calculated efficiently using

\begin{equation}
    D^2_w(f, \tilde{f}) = \bm{\delta}' W(\omega)J W(\omega) \bm{\delta}.
\end{equation}
Beforehand, the same formula was derived without the $W(\omega)$-matrix in equation (\ref{eq:Jmatrix}).  $W(\omega)$
is a diagonal matrix of size $K$. The parameter $\omega$ is a (potentially multidimensional)
parameter describing the weighting. This parameter is estimated during the maximum likelihood
optimization. One possibility would be to include each diagonal entry of $W$ into $\omega$ and just
restrict it to be positive. This is unfortunate for two reasons. First this potentially increases
the number of parameters to be estimated by ML dramatically. Secondly, the GP model would no
longer be uniquely identifiable. The covariance parameter $\theta$ and the weighting parameter
$\omega$ could be exchanged without changing the model. To overcome the identification problem, the
entries of $W$ are restricted to be nonnegative and to sum up to 1: $W_{ii} \geq 0, tr(W) = 1$. In
order to reduce the number of parameters, here a parametric description of the weighting by a beta
distribution is used. The beta distribution is a very flexible distribution with support $[0,
1]$. It is described by two parameters, which both need to be greater than 0. Let $dbeta(t, \omega)$
the density of a beta distribution with parameters $\omega$. Then the weighting matrix is defined as
\begin{equation}
    \tilde{W}_{ii} = dbeta(t_{imax}, \omega), ~~~~~W(\omega) := \tilde{W}(\omega) / tr(\tilde{W}(\omega)).
\end{equation}
The value $t_{imax}$ is the argument value, where the $i$th basis spline has its maximum, i.e. the place where the $i$th basis has the highest influence. As the two parameters in $\omega$ are just restricted to be $\geq 0$, the numerical optimization of these two parameters can easily be incorporated into the ML estimation procedure for the covariance parameters.

\section{Application}
\label{sec:application}
\subsubsection{Theoretical example 1}
In order to check if the estimation of the covariance parameters work comparable both for real number inputs and functional inputs the following example is used:
\begin{align*}
g_1(x, f) =& x_1 + 2 x_2 +  4\int_0^1 t f_1(t)dt + 1  \int_0^1 f_2(t)dt\\
\end{align*}
The first real number and the second functional input are of the same importance, i.e. the function $x_1$ and the function $\int_0^1{f_2(t)}dt$ have the same output domain. At the same time, the second real number input and the first functional input are comparably influential but are more important than the first real number and functional input. 

\begin{figure}[ht]
  \centering
  \includegraphics[width=\textwidth]{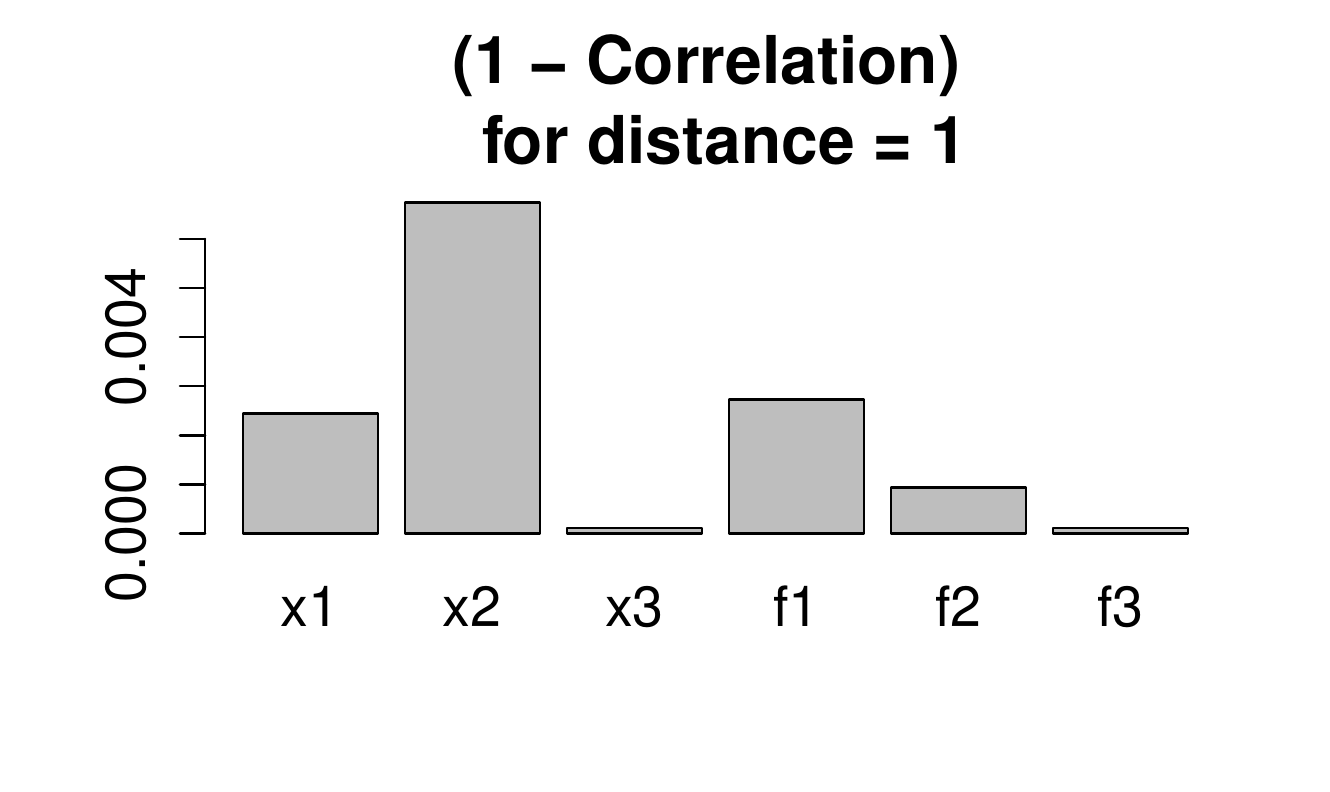}
\caption{Sensitivity plot for the first theoretical example using a GP model including weighting.}
\label{fig:Sensitivityplotfampli}
\end{figure}

\begin{figure}[ht]
  \centering
  \includegraphics[width=\textwidth]{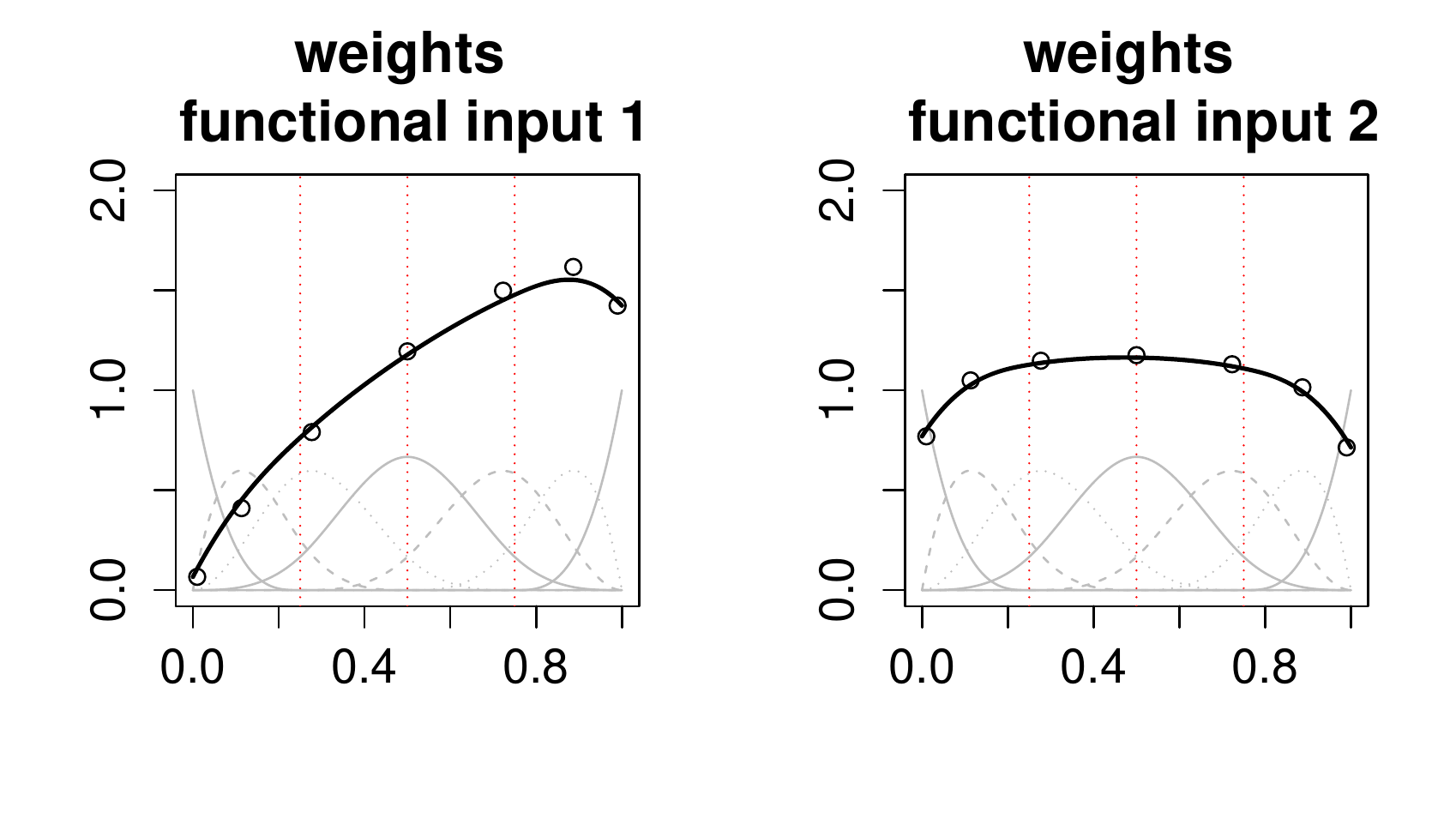}
\caption{Weighting plot for the first theoretical example. As the third functional input is (correctly) rated as unimportant, only the first two weighting plots are shown.}
\label{fig:weightingplotfampli}
\end{figure}

A DoE with 20 runs and 3 real number and 3 functional inputs is set up (hence there are inactive input parameters). The corresponding B-splines have 7 basis functions and are of order 4. Afterwards a GP model with the Matern5/2 covariance kernel inlcuding weighting is fitted to the data. 
The result of the covariance plot is shown in Figure \ref{fig:Sensitivityplotfampli} and the weighting plots are shown in Figure \ref{fig:weightingplotfampli}. 

\subsubsection{Theoretical example 2}
\label{sec:example2}
The second example has again 3 scalar inputs and 3 functional inputs $f_k (t) \in C^0( [0, 1 ]), k = 1, 2, 3$, satisfying boundary constraints $0 \leq f_k \leq 1$, but this time the function is chosen to be more complex. The real number part is the well known Branin function (\cite{DixonSzego1978Branin}) plus one inactive input and the functional part of the example has 2 active inputs and one inactive, including interactions between the real number and functional inputs. 

\begin{align*}
g_2(x, f) =&  (x_2 - \frac{5}{4 \pi^2 } x_1^2 + \frac{5}{\pi} x_1 - 6)^2 + 10(1 - \frac{1}{8 \pi}) \text{cos}(x_1) + 10  \\
        + & \frac{4}{3}\pi \left( 42 \int_{-5}^{10} f_1(t)(1 - t) dt + \pi  ((x_1 + 5) / 5 + 15) \int_0^1 t f_2(t)dt  \right).
\end{align*}

\begin{figure}[ht]
  \centering
  \includegraphics[width=\textwidth]{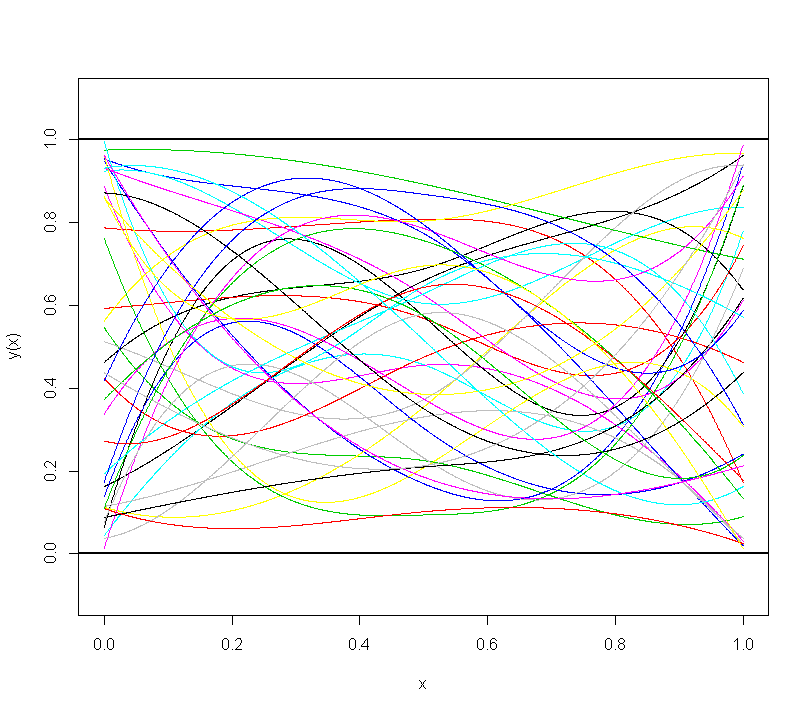}
\caption{Candidate set with $n = 40, K = 5, m = 4$.}
\label{fig:candidatesettheoreticalexample}
\end{figure}

In order to construct a design, the strategy described above is applied with $n = 40, K = 7, m = 4$.
The candidate set is shown in figure \ref{fig:candidatesettheoreticalexample}.

\begin{figure}[ht]
  \centering
  \includegraphics[width=\textwidth]{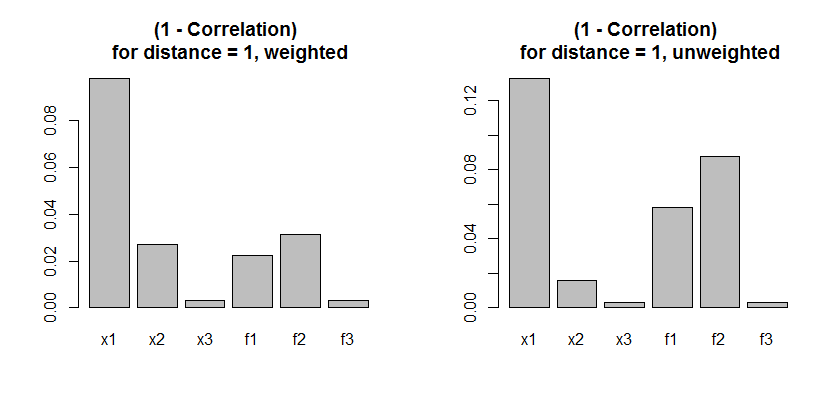}
\caption{Sensitivity plots based on the covariance parameters for the two models including weighting (left hand side) and without weighting (right hand side).}
\label{fig:Sensitivityplot}
\end{figure}

\begin{figure}[ht]
  \centering
  \includegraphics[width=\textwidth]{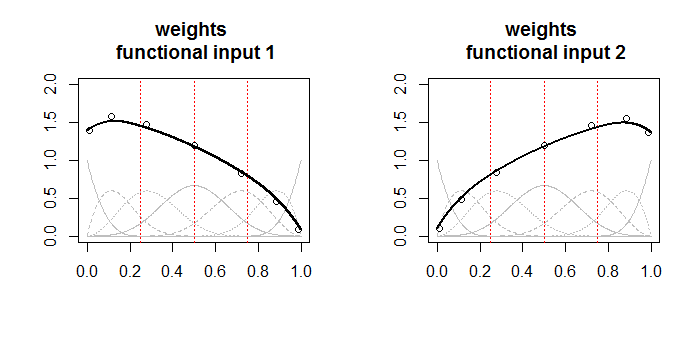}
\caption{Weighting plot for the GP model including weighting. Areas on the x-scale with a high value are of higher importance on the output than areas with smaller values.}
\label{fig:Weightingplot}
\end{figure}

A generalized LHD according to the methodology described above has been constructed and two GP models have been estimated: One without any weighting for the functional inputs and a second one including weighting for the functional inputs as described in the last chapter. As a covariance kernel, the Matern $5/2$ kernel has been used. Both GP models have been used in order to make predictions for 300 randomly selected sets of inputs points and input functions in order to validate the prediction quality of the two models. For both models, the weighted and the unweighted one, the covariance parameters are summarized in a bar plot in order to illustrate, which inputs are important. In this bar plot, $1 - g_k(1;\theta_k)$ is plotted, where $g_k$ is the kernel of the covariance function chosen (see Figure (\ref{fig:Sensitivityplot})). For the weighted model, the result of the weighting procedure is plotted in figure (\ref{fig:Weightingplot}). Here again $1 - g_k(1;\theta_k)$ is plotted in a bar plot. Furthermore, a Bspline is plotted, where the weights obtained from the maximum likelihood procedure are used as $\beta$-coefficients. This plot indicates which part of the support has high importance and which part has low importance.

\begin{figure}[ht]
  \centering
  \includegraphics[width=\textwidth]{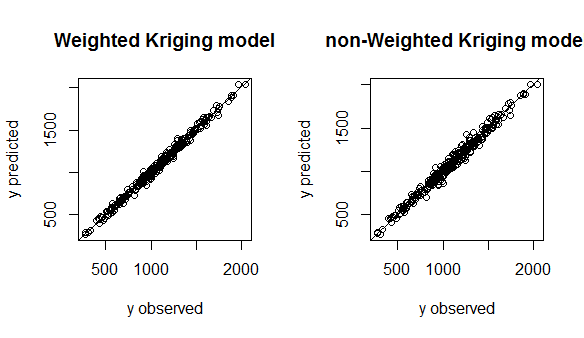}
\caption{Prediction plots for the two models including weighting (left hand side) and without weighting (right hand side).}
\label{fig:predictionplots}
\end{figure}

Both models deliver a good prediction based on prediction plots (see figure (\ref{fig:predictionplots})), where the weighted version has slightly better RMSE (0.055) than the nonweighted version (0.081) based on 300 independent observations of the theoretical example.

\subsection{Springback analysis}

In deep drawing sheet metal forming, the final shape of a part depends on the elastic energy stored during the
process of the forming. The energy is influenced by a number of process parameters like blankholder
force and friction. Springback, one of the main sources of geometrical inaccuracy, can be
predicted by these parameters in simulation models. Usually, the analysis is limited to constant input parameters. The goal here is to achieve better predictions and deeper information to the springback development by varying the process parameters blankholder force and friction in time using the
norm-based function analysis approach. 

An explicit Finite Element Method (FEM) via LS-DYNA is used which takes two parameters
as input, the friction coefficient ($f_{F}$) and the blankholder force ($f_{B}$), which can be
varied externally during the punch travel.

\begin{figure}[ht]
  \centering
  \includegraphics[width=\textwidth]{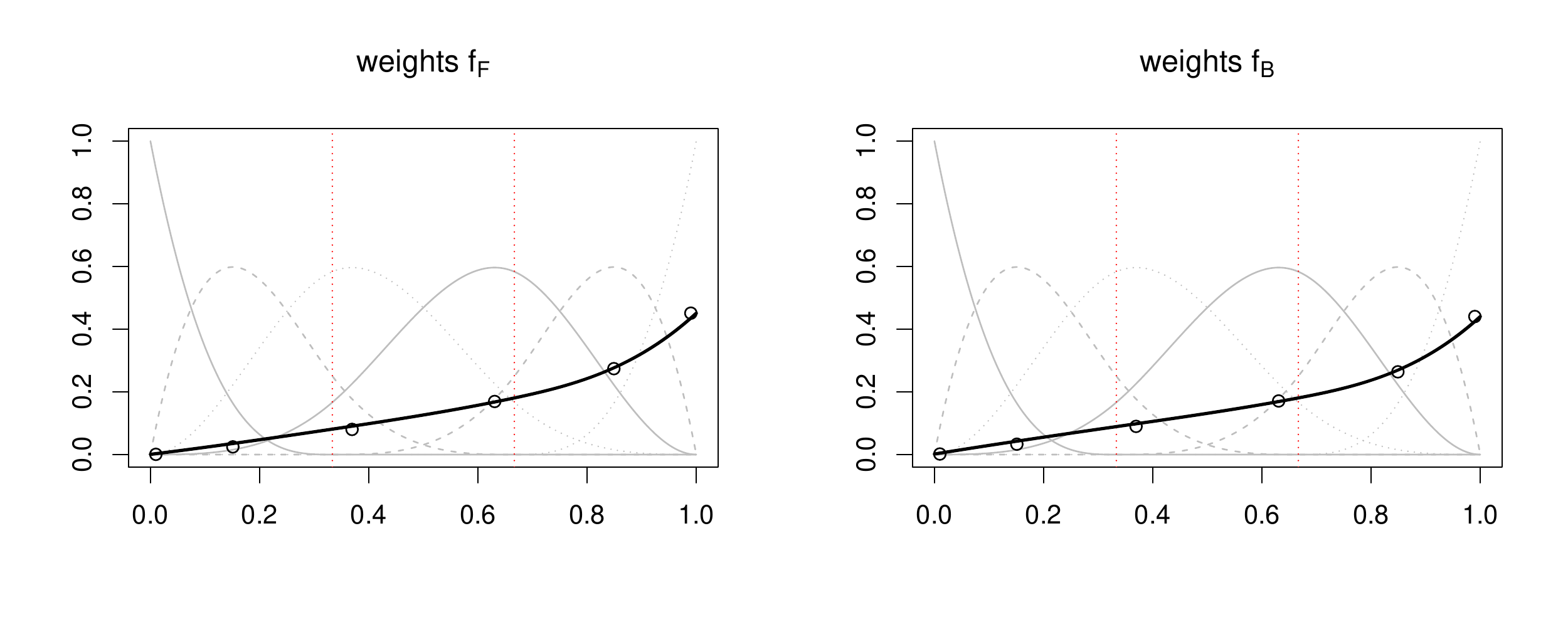}
\caption{Weighting plots for the springback FEM model.}
\label{fig:springback-weighting}
\end{figure}

A generalized LHD with 40 runs, using 6 B-spline basis functions of order 4 is constructed and
performed in the FEM model. On these data, a functional Kriging model including weighting on a Gauss
covariance kernel is fitted. The value $1 - g_k(1;\theta_k)$ is around 0.78 for $f_{F}$ and 0.28 for
$f_{B}$ indicating a much larger influence for friction as for blankholder force on the
springback. A weighting plot as in Fig. \ref{fig:Weightingplot} can be seen in
Fig. \ref{fig:springback-weighting}. It can be found that in the FEM model the effect of the inputs
on the springback reduction is increases towards the end of the punch travel. This shows the
importance of careful settings at the end of the process, where the flange of the part is formed.

\subsection{Exploration of B-splines}

\begin{figure}[ht]
  \centering
  \includegraphics[width=\textwidth]{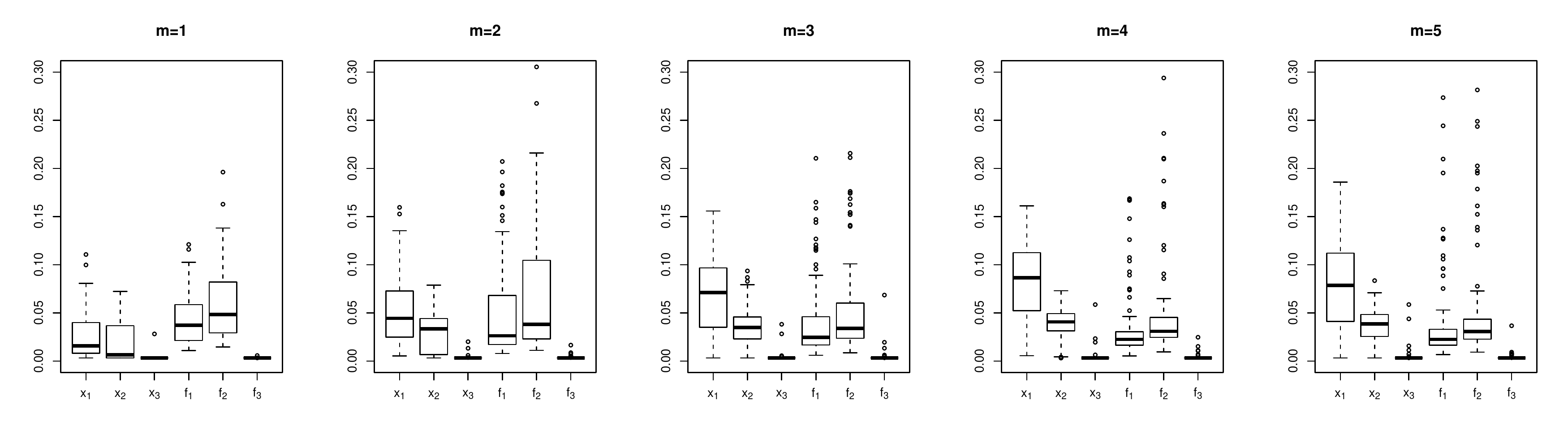}
\caption{B-spline order comparison: Boxplots of the values $1 - g_k(1;\theta_k)$.}
\label{fig:comparison}
\end{figure}

We revisit the theoretical example 2 of Section \ref{sec:example2} for a small study in which we
examine the effect of the B-spline order $m$ to the presented functional design and modelling
approach. We compare five different orders, $1,2,3,4$ and $5$. For each order we set up a design of
size $n=20$ with $K=7$ basis functions and construct a surrogate model. The constant number $K$
ensures a comparable number of model parameters between the different orders. The procedure is
repeated 100 times for each order and 5 test data sets of size 600, one for each order, are set up
for comparison. Table \ref{tab:comparison} shows the resulting root mean square errors (RMSE),
averaged over the 100 models. The spline order $m=4$ performs best here. We conclude that, at least
in our experience, the B-spline order 4 can be recommended. Figure \ref{fig:comparison} shows
boxplots of the values $1 - g_k(1;\theta_k)$, comparable to Figure (\ref{fig:Sensitivityplot}). The
box sizes give an impression of the accuracy of the covariance estimates.

\begin{table}[ht]
\centering
\begin{tabular}{lrrrrr}
  \hline
$m$ & 1 & 2 & 3 & 4 & 5 \\
  \hline
Average RMSE &  56.29 & 53.10 & 44.81 & 37.07 & 40.065\\
   \hline
\end{tabular}
\caption{B-spline order comparison: Average RMSE values between the predicted and true values of the 5 test data sets.}
\label{tab:comparison}
\end{table}

\section{Conclusion}
\label{sec:conc}

In this article a methodology for incorporating functional inputs and scalar inputs into simulation experiments via the use of B-splines is presented. Therefore designs and metamodels are developed.
For constructing a space-filling designs, a distance-based approach is presented, which works in two steps. In a first step a candidate set for the functional input parameters is constructed and in the second step a design for the functional as well for the scalar inputs is constructed in a Latin hypercube manner.
Given scalar outputs from a simulation, the data can be modelled by a GP model and the covariance parameters are used in order to rank the inputs by importance.
In order to learn more about the behaviour of the functional inputs, a weighting process can be introduced, which can analyze, where a functional input is of high importance. This gives an attractive possibility to learn more about the behaviour of the functional inputs. But this benefit comes with the cost of introducing additional parameters and therefore with a more demanding optimization process.
The weighting process is not so beneficial for improving prediction but it aims at learning more about the functional inputs. Although when the data set is large enough the prediction for the weighted GP model has often been slightly better than for the unweighted version, especially for small sample sizes, the estimation process of the parameters for weighted GP model does not work as reliably as for the non-weighted GP model.
All in all, the methodology developed incorporates functional inputs in a way that the functional character is not changed but still computations are feasible. Fundamental to this has been the usage of functional norms in order to incorporate functional inputs and the usage of B-splines as a representation of functional inputs.

Acknowledgements:

Andon Iyassu, for his help on editing.

\bibliography{LitMuehlenstaedt}

\begin{thebibliography}{}

\bibitem[{Bayarri} et~al., 2007]{BayarriEtc2007ValidationFunOut}
{Bayarri}, M.~J., {Berger}, J.~O., {Cafeo}, J., {Garcia-Donato}, G., {Liu}, F.,
  {Palomo}, J., {Parthasarathy}, R.~J., {Paulo}, R., {Sacks}, J., and {Walsh},
  D. (2007).
\newblock {Computer Model Validation with Functional Output}.
\newblock {\em Annals of Statistics}, 35:1874 -- 1906.

\bibitem[Currin et~al., 1991]{CurMitMor91baye}
Currin, C., Mitchell, T., Morris, M., and Ylvisaker, D. (1991).
\newblock Bayesian prediction of deterministic functions, with applications to
  the design and analysis of computer experiments.
\newblock {\em Journal of the American Statistical Association},
  86(416):953--963.

\bibitem[de~Boor, 2001]{deBoor2001GuideToSplines}
de~Boor, C. (2001).
\newblock {\em A practical guide to splines}.
\newblock Springer, New York.

\bibitem[{Dixon} and {Szego}, 1978]{DixonSzego1978Branin}
{Dixon}, L.~C.~W. and {Szego}, G.~P. (1978).
\newblock {The global optimization problem: An introduction}.
\newblock {\em Towards global optimization}, 2:1 -- 15.

\bibitem[Fang et~al., 2006]{FanLiSud06desi}
Fang, K.-T., Li, R., and Sudjianto, A. (2006).
\newblock {\em Design and Modeling for Computer Experiments}.
\newblock Computer Science and Data Analysis Series. Chapman \& Hall/CRC, New
  York.

\bibitem[Jones et~al., 1998]{JonSchonWelch98EGO}
Jones, D., Schonlau, M., and Welch, W. (1998).
\newblock Efficient global optimization of expensive black-box functions.
\newblock {\em Journal of Global Optimization}, 13:455--492.

\bibitem[McKay et~al., 1979]{MckBecCon79Comp}
McKay, M., Beckman, R., and Conover, W. (1979).
\newblock A comparison of three methods for selecting values of input variables
  in the analysis of output from a computer code.
\newblock {\em Technometrics}, 21(2):239--245.

\bibitem[Morris, 2012]{Morris2012KrigingWithTimeVaryingInputs}
Morris, M. (2012).
\newblock Gaussian surrogates for computer models with time-varying inputs and
  ouptputs.
\newblock {\em Technometrics}, 54:42--50.

\bibitem[Morris, 2014]{Morris2014MaximinDoE}
Morris, M. (2014).
\newblock Maximin distance optimal designs for computer experiments with
  time-varying inputs and outputs.
\newblock {\em Journal of statistical Planning and Inference}, 144:63--68.

\bibitem[Morris and Mitchell, 1995]{MorMit95expl}
Morris, M. and Mitchell, T. (1995).
\newblock Exploratory designs for computational experiments.
\newblock {\em Journal of Statistical Planning and Inference}, 43:381--402.

\bibitem[{R Core Team}, 2013]{RCoreTeamR}
{R Core Team} (2013).
\newblock {\em R: A Language and Environment for Statistical Computing}.
\newblock R Foundation for Statistical Computing, Vienna, Austria.
\newblock {ISBN} 3-900051-07-0.

\bibitem[Ramsay and Silverman, 1997]{RamSil97func}
Ramsay, J. and Silverman, B. (1997).
\newblock {\em Functional Data Analysis}.
\newblock Springer, New York.

\bibitem[Rasmussen and Williams, 2006]{GPML}
Rasmussen, C. and Williams, C. (2006).
\newblock {\em Gaussian Processes for Machine Learning}.
\newblock the MIT Press.

\bibitem[Roustant et~al., 2012]{RouGinDevDiceKriging}
Roustant, O., Ginsbourger, D., and Deville, Y. (2012).
\newblock {DiceKriging}, {DiceOptim}: Two {R} packages for the analysis of
  computer experiments by kriging-based metamodeling and optimization.
\newblock {\em Journal of Statistical Software}, 51(1):1--55.

\bibitem[Sacks et~al., 1989a]{SacSchWel89desi}
Sacks, J., Schiller, S., and Welch, W. (1989a).
\newblock Design for computer experiments.
\newblock {\em Technometrics}, 31:41--47.

\bibitem[Sacks et~al., 1989b]{SacWelMit89desi}
Sacks, J., Welch, W., Mitchell, T., and Wynn, H. (1989b).
\newblock Design and analysis of computer experiments.
\newblock {\em Statistical Science}, 4:409--435.

\bibitem[Santner et~al., 2003]{SanWilNot03desi}
Santner, T., Williams, B., and Notz, W. (2003).
\newblock {\em The Design and Analysis of Computer Experiments}.
\newblock Springer Series in Statistics. Springer Verlag, New York.

\bibitem[Shi and Shoi, 2011]{ShiShoi2011GPFunctionalDataBook}
Shi, J.~Q. and Shoi, T. (2011).
\newblock {\em Gaussian Process Regression Analysis for Functional Data}.
\newblock Chapman \& Hall.

\bibitem[Shi et~al., 2007]{ShiWanMurTit2007GPFRforBatchData}
Shi, J.~Q., Wang, B., Murray-Smith, R., and Titterington, D.~M. (2007).
\newblock Gaussian process functional regression modeling for batch data.
\newblock {\em Biometrics}, 63(3):714--723.

\end{thebibliography}
\bibliographystyle{apalike}

\end{document}